# An optical all-pass filter realized by using self-compensation of loss

Lu Xu[1,2], Yuan Yu[1,2]*, Xiaolong Liu[1], Xuewen Shu[1] & Xinliang Zhang[1]*

**Abstract:** Filters are key devices in optical systems and are usually applied to signal suppression or selection based on their amplitude frequency responses. Different from amplitude filtering, an all-pass filter (APF) is a unique type of filter that exhibits a uniform amplitude and variable phase in the frequency response. Traditionally, an APF is designed to be ideal and lossless, which is impossible in practice. Here we demonstrate a method of realizing an APF even when transmission loss exists, and a possible structure for realizing the lossy APF is proposed. The method indicates that a lossy APF can be realized by introducing self-compensation of loss to the structure of an ideal single-stage APF, and an APF based on silicon-on-insulator (SOI) foundry is fabricated and demonstrated. The method of self-compensation of loss can also be used to realize APFs in the electrical domain. We anticipate that our work will inspire realizations of more APF-related devices that once seemed impossible in practice.

Filters have been widely used in signal processing systems to select desired signals and suppress interferences and noise based on their amplitude frequency responses[1-5]. Instead of manipulating power, an all-pass filter (APF) aims to manipulate the signal phase and exhibits a constant amplitude response with respect to frequency. In an APF, the phase response can be designed for different applications by utilizing suitable structures. With this unique characteristic, APFs are highly desired for occasions where only a phase variation is needed, such as the phase shifters[6,7], dispersion compensation[8-10], time delays[11,12], ultrafast all-optical clock recovery[13] and Hilbert transformers[14,15].

General optical APF structures have been achieved in ideal lossless circumstances[16]. The research shows that in a lossless $2\times2$ network structure that consists of two inputs and two outputs, by connecting one output to one input with a delay line, the resulting structure is a single-stage APF. However, lossless transmission is impossible in practical waveguides, and the implementations of an APF in this manner are actually approximations of ideal APFs and are not real APFs[9-12]. The same problem also exists in the theoretical models[17-19] and fabricated devices[20-23] of electrical APFs. Up to now, to the best of our knowledge, an APF has not yet been realized with lossy transmission.

To realize an APF, precise compensation of the transmission loss in the waveguide is of fundamental importance. In active waveguides, the transmission loss can be compensated by controlling and tailoring the gain spectrum in theory. However, it is rather difficult to precisely equalize the gain and loss, because the equalization is very sensitive to fluctuations in the input power and the injection current of the active waveguide. Therefore, to avoid lasing, the gain must be set to be lower than the loss, and consequently, the amplitude variations in the amplitude frequency response cannot be eliminated.

[1]Wuhan National Laboratory for Optoelectronics (WNLO), Huazhong University of Science and Technology (HUST), Wuhan, China. [2]These authors contributed equally: Lu Xu, Yuan Yu. Correspondence and requests for materials should be addressed to Y.Y. (email: yuan_yu@hust.edu.cn) or to X.Z. (email: xlzhang@mail.hust.edu.cn)

To resolve the challenge of precise loss compensation, here we propose to compensate for transmission loss in passive waveguides by using a novel method called self-compensation of loss, in which the compensation is a power scaling copy of the input. A feasible source for this kind of compensation is to take a slice from the input. When the compensation precisely equals to the transmission loss, an APF can then be realized.

Here, we consider an all-pass microring resonator (MRR), which exhibits all-pass transmission only for lossless transmission (see Methods), as an example to demonstrate the method. A detailed demonstration of the realization of a general single-stage lossy APF using self-compensation of loss is given in Methods. The transfer function of a lossy all-pass MRR (Fig. 1a) can be expressed as

$$H = \frac{r_1 - ae^{i\varphi}}{1 - r_1 ae^{i\varphi}} \tag{1}$$

where $r_1$ is the self-coupling coefficient, $\varphi$ is the round-trip phase shift of the MRR and $a$ is the round-trip amplitude transmission, including both the propagation loss in the MRR and the coupling loss in the coupler[24] (the influence of the coupling loss on the transmission of the APF is discussed separately in Methods). Here, $a=exp(-\alpha L/2)$, where $\alpha$ is the power attenuation coefficient and $L$ is the perimeter of the MRR. We can conclude that $a=1$ for lossless transmission. Based on a lossy all-pass MRR, the structure of the APF with loss compensation (Fig. 1b) is the same as that of an add-drop MRR, and the compensation is injected into the MRR via the add port (port 3 in Fig. 1b). It should be stressed that the phase of the compensation is the same as that of the input.

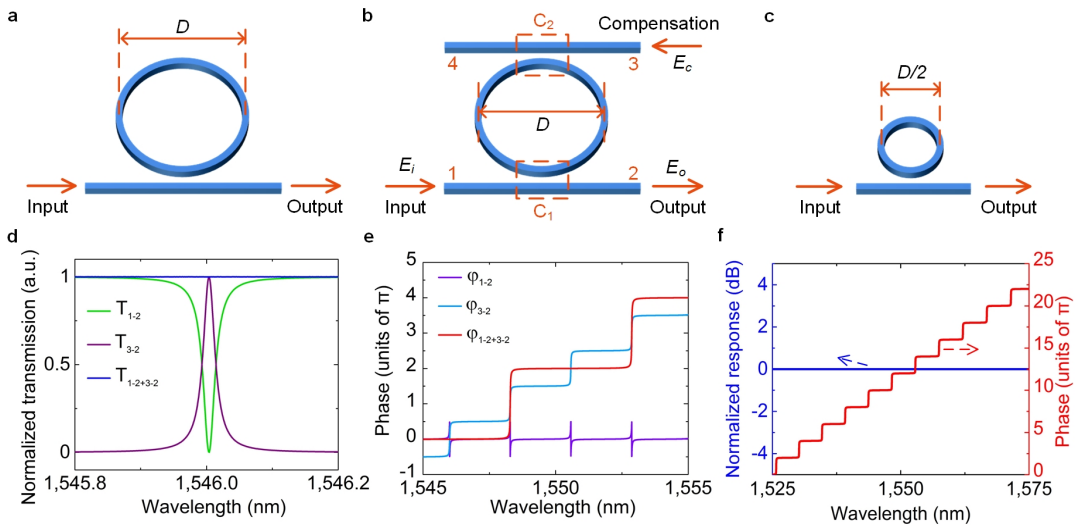

**Fig. 1.** The MRR-based APF. **a**, The structure of a lossy all-pass MRR. The diameter of the MRR is represented by $D$. **b**, The structure of the all-pass MRR with self-

compensation of loss. The structure is the same as that of an add-drop MRR. Ports 1, 2, 3 and 4 represent the input, through, add and drop ports of the add-drop MRR, respectively. $C_1$ and $C_2$ represent the two directional couplers connecting ports 1 and 2 and ports 3 and 4, respectively. **c**, An ideal lossless all-pass MRR. The diameter of the MRR is $D/2$. **d**, **e**, The transmission spectra (**d**) and the phase responses (**e**) of the add-drop MRR when only the input ($T_{1-2}$ and $\varphi_{1-2}$), only the compensation ($T_{3-2}$ and $\varphi_{3-2}$), and the combined input and compensation ($T_{1-2+3-2}$ and $\varphi_{1-2+3-2}$) are injected into the device, respectively. **f**, Simulated results of the MRR-based APF in **b**.

To realize the APF, two conditions must be satisfied (see Methods). Here, in the case of an MRR-based APF, condition 1 is that the add-drop MRR is critically coupled[24] for the input at port 1, indicating that $r_1=r_2a$, where $r_1$ and $r_2$ are the self-coupling coefficients of couplers $C_1$ and $C_2$ (Fig. 1b), respectively. Assume that the amplitudes of the input at port 1 and the compensation at port 3 are represented by $E_i$ and $E_c$, respectively. Condition 2 is that the amplitude ratio between $E_c$ and $E_i$, which is represented by $x$, satisfies $x=k_1/(k_2\sqrt{a})$ (see Methods). Under the two conditions, when the input and the compensation are injected into the MRR alternatively, the corresponding transmission spectra from port 1 to port 2 and from port 3 to port 2 are shown by the green and purple curves in Fig. 1d, respectively. It can be observed that the two transmission spectra are complementary to each other. However, when the two conditions are satisfied and the input and compensation are injected into the MRR simultaneously, the transfer function of the add-drop MRR from port 1 to port 2 is calculated to be (see Methods)

$$H = \frac{E_o}{E_i} = \frac{r_1 - e^{i(\varphi/2)}}{1 - r_1 e^{i(\varphi/2)}} \qquad (2)$$

where $E_o$ represents the amplitude of the output at port 2 (Fig. 1b). By comparing equations (1) and (2), we can observe that the add-drop MRR is equivalent to a lossless all-pass MRR, which is an ideal APF. From equation (2), it can be calculated that the transfer function satisfies $|H|^2=1$, shown as the blue curve in Fig. 1d. This is in accordance with the definition of an APF. The corresponding phase responses (Fig. 1e) show that, when both the input and the compensation are injected into the add-drop MRR simultaneously, the free spectral range (FSR) of the phase response (red curve in Fig. 1e) is twice the FSR of the phase response when only the input or the compensation is injected into the MRR, indicating that the round-trip phase shift of the equivalent lossless all-pass MRR is $\varphi/2$. Consequently, under the two conditions, the lossy add-drop MRR after loss compensation is equivalent to the ideal lossless APF shown in Fig. 1c. In the equivalent ideal lossless APF, the self-coupling and cross-coupling coefficients are $r_1$ and $k_1$, respectively, and the perimeter is one-half of that of the lossy add-drop MRR (Fig. 1b). Fig. 1f shows the simulated response of the lossy add-drop MRR when the optical signal is transmitted from port 1 to port 2 after self-compensation of loss, which is equivalent to that of the ideal lossless all-pass MRR (Fig. 1c). In Fig. 1f, we can observe that the amplitude response is constant with respect to wavelength while the phase varies periodically. Therefore, the combined result of conditions 1 and 2 is that the transmission of the input in the MRR is exactly compensated by the compensation, and a real APF can be achieved.

To eliminate the influence of fabrication error on the deviations of the coupling coefficients, we propose a Mach-Zehnder interferometer (MZI)-assisted MRR structure (Fig. 2a) to realize the APF. In this structure, the MZIs are designed to achieve the desired coupling coefficients with controllable adjustments[25]. $MZI_1$ and $MZI_2$ each have two directional couplers and the self-coupling and cross-coupling coefficients of the couplers are both $r_1$ and $k_1$ in $MZI_1$ and are both $r_2$ and $k_2$ in $MZI_2$, respectively. In Figs 2a and b, the centres of the couplers are marked with orange dots and represented by capital letters. Then, the waveguides can be marked by capital letters at the terminals of the waveguides. The lengths of the waveguides AC, BD, EG and FH are all $L_1$, and their amplitude transmission $a_1$ satisfies $a_1=exp(-\alpha L_1/2)$. Similarly, the lengths of the delay lines DE and GB are both $L_0$, and their amplitude transmission $a_0$ satisfies $a_0=exp(-\alpha L_0/2)$. $\varphi_0$ is the phase shift of the delay lines DE and GB, $\varphi_1$ is the phase shift of AC, BD and EG and $\varphi_2$ is the phase shift of FH. According to conditions 1 and 2, it can be demonstrated that the structure in Fig. 2a is an APF when the parameters satisfy (see Methods)

$$r_2 = \sqrt{\frac{r_1^2 - k_1^2 + y a_0^2 a_1^2}{(y+1) a_0^2 a_1^2}}$$
$$x = \frac{2 k_1 r_1}{(y+1) k_2 r_2 a_0 a_1}$$
(3)

where $y = exp[i(\varphi_2 - \varphi_1)]$ and $x = E_c/E_i$. Fig. 2b shows the structure of an ideal lossless APF based on an MZI (see Methods). The phase shifts of A'C' and B'D' are both $\varphi_1$, which is the same as the phase shifts of AC and BD. The phase shift of D'I'B' is the same as that of DE or GB and equal to $\varphi_0$. The waveguides in Fig. 2b are all lossless. It can be demonstrated that the MZI-assisted MRR in Fig. 2a is equivalent to the ideal lossless APF in Fig. 2b (see Methods).

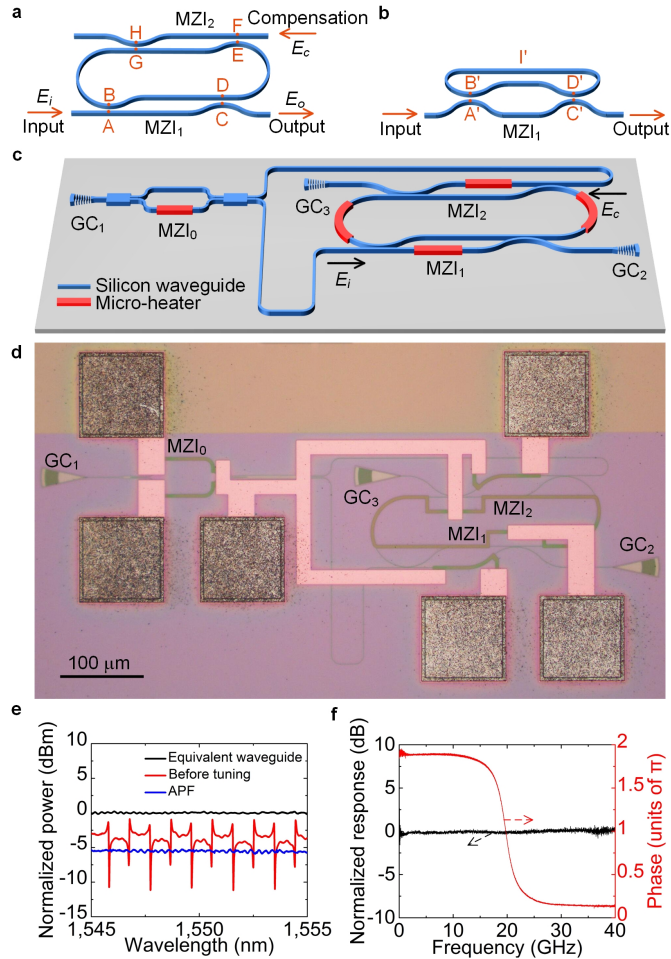

**Fig. 2.** Realization of an APF based on the MZI-assisted MRR. **a**, The structure of the MZI-assisted MRR. **b**, The structure of an ideal lossless APF based on an MZI. **c**, The structure of the designed APF. **d**, A micrograph of the fabricated APF based on silicon-on-insulator (SOI) foundry. Here, a spare micro-heater is added to the upper arm of $MZI_0$. **e**, Measured optical transmission spectra of the fabricated device. The blue curve shows the measured optical spectrum of the fabricated APF, the red curve shows the measured optical spectrum of the device when no voltages are applied to the micro-heaters, and the black curve shows the transmission of the equivalent optical waveguide. **f**, Amplitude and phase frequency responses of the proposed APF measured by a vector network analyzer (VNA).

According to the theoretical analysis, the structure shown in Fig. 2c can be designed to realize an APF based on the MZI-assisted MRR by using self-compensation of loss. The optical signal is input into the photonic chip via a grating coupler ($GC_1$ in Fig. 2c) and equally split into the two arms of $MZI_0$ by a 1×2 multimode interferometer (MMI)[26]. $MZI_0$ is then connected with a 2×2 MMI, and the power splitting ratio between the two outputs can be adjusted by changing the phase difference between the two arms of $MZI_0$[27] via a micro-heater deposited on the lower arms of $MZI_0$. After $MZI_0$, the optical signal at the lower port acts as the input $E_i$, and the optical signal at the upper port acts as the compensation $E_c$. Then, two waveguides of the same length (ensuring that no phase difference exists between the input and compensation) are used to deliver the input and compensation to the MRR, respectively. Based on the designed lossy APF shown in Fig. 2c, the fabricated APF based on silicon-on-insulator (SOI) foundry is shown in Fig. 2d.

In the structure shown in Fig. 2c, conditions 1 and 2 can be satisfied by adjusting the voltages applied to the micro-heaters. When the two conditions are not satisfied simultaneously, the optical power transmission from $GC_1$ to $GC_2$ varies with respect to wavelength (red curve in Fig. 2e). By adjusting the voltages applied to the micro-heaters to satisfy conditions 1 and 2, the power transmission spectrum at the output of the APF is shown by the blue curve in Fig. 2e. Compared with the red curve, we can observe that the periodic power variations at the resonance wavelengths no longer exist. We also compare the measured transmission spectrum of the APF with that of an equivalent single-mode waveguide (see Methods) shown as the black curve in Fig. 2e. We can observe that the transmission spectrum of the APF is in good agreement with that of the equivalent optical waveguide except for an extra transmission loss of approximately 5 dB, which is caused by the compensation used for self-compensation of loss. In the

APF operation, one part of the input optical signal is split into the upper waveguide by $MZI_0$ and used as the compensation. Therefore, extra transmission loss is generated. Slight ripples with a maximal amplitude of approximately 0.3 dB and an FSR of approximately 0.3 nm are generated by the Fabry-Pérot cavities formed by $GC_1$ and $GC_2$ and by $GC_1$ and $GC_3$, simultaneously. These ripples can be reduced by further optimizing the grating couplers. To measure the amplitude and phase frequency responses precisely, the fabricated APF is also measured by a vector network analyzer (VNA) (see Methods), and the results are shown in Fig. 2f. We can observe that the amplitude remains constant while the phase has a variation of 1.8 π within 40 GHz. Therefore, we can conclude that the APF is successfully realized. It should be noted that all-pass transmission can alternatively be achieved via $GC_3$ when the micro-heaters are adjusted correspondingly to satisfy the two conditions for the optical transmission from $GC_1$ to $GC_3$.

Since only phase variations exist in the frequency response of the APF and the amplitude remains constant, some interesting applications can be explored based on the fabricated APF (the experimental details are shown in Methods). The APF is an ideal device for a microwave photonic phase shifter. By adjusting the voltage applied to the micro-heaters deposited on the MRR, the measured microwave phase shifts and the corresponding amplitudes with respect to frequency are shown in Figs 3a and b, respectively. In Fig. 3b, it can be observed that there is a power variation of approximately 0.4 dB in every amplitude frequency response curve, which is generated by the transmission ripples caused by the Fabry-Pérot effect (Fig. 2e). During the adjustment of the microwave phase shift, we can see that the output microwave power changes slightly and the power variation is less than 0.8 dB. This is caused by thermal crosstalk during the adjustment of the micro-heater on the MRR, which can be reduced by using thermal isolation trenches[28].

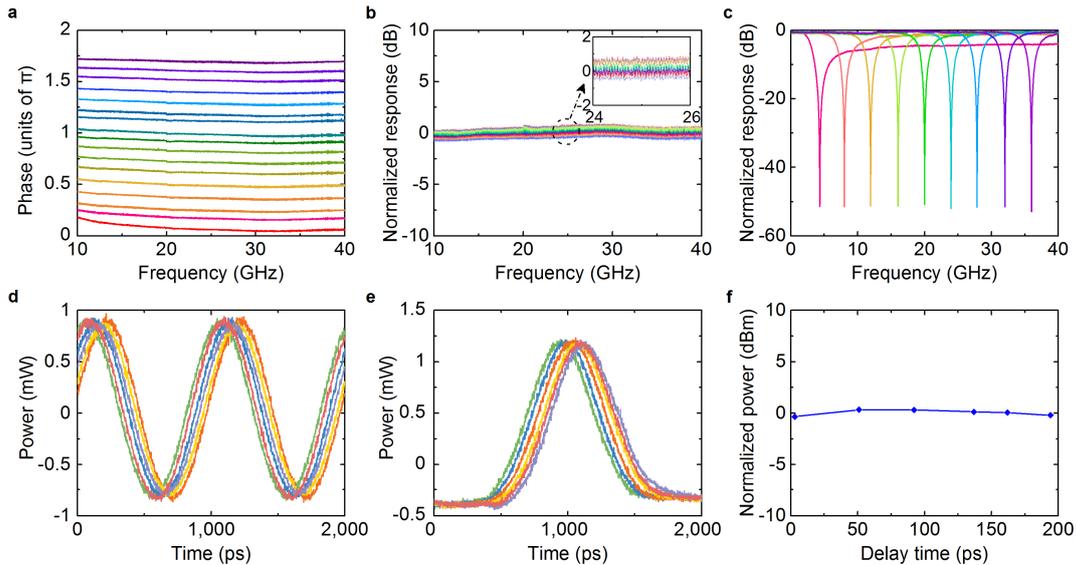

**Fig. 3.** Results of possible applications based on the fabricated APF. **a**, **b**, Phase (**a**) and amplitude (**b**) frequency responses with different phase shifts when the APF works as a microwave photonic phase shifter. **c**, The microwave photonic notch filter based on the APF with different central frequencies. **d**, **e**, Different measured time

delays of the APF when the input is a sinusoidal signal with a frequency of 1 GHz (**d**) and a Gaussian pulse with a bandwidth of 500 ps and a repetition rate of 500 MHz (**e**). **f**, The normalized output optical power for different time delays.

Based on the optical phase shifter, a single notch microwave photonic filter is also realized (Fig. 3c). The rejection ratio is over 50 dB, and the central frequency can be tuned from 0 to 40 GHz, which is limited by the bandwidth of the modulator. Based on the proposed APF, tunable time delays can also be achieved with almost invariant output power. A sinusoidal signal and a Gaussian pulse are used in the experiment to measure the time delay, respectively. The measured maximum time delays of both the sinusoidal signal and Gaussian pulse are approximately 200 ps (Figs 3d and e). When the time delay is changed by adjusting the voltage applied to the micro-heater on the MRR, the measured power variation of the optical signal is less than 0.4 dB (Fig. 3f).

In summary, by using self-compensation of loss, we have demonstrated a real APF based on an add-drop MRR even when transmission loss exists, and an APF based on an MZI-assisted MRR has been demonstrated and fabricated using the SOI foundry. Applications of the proposed APF in a microwave phase shifter, a microwave photonic filter, and time delays have been explored. We believe that the method of self-compensation of loss employed for realizing the APF holds great potential in numerous fields of research. Our work opens up a possible path for realizing devices that seemed to be impossible in the past, such as amplitude distortion-free Hilbert transformers[15,29,30] and microwave photonic phase shifters. With fabrication improvements in the future, the realization of an APF will be simplified, and more efforts should be devoted to exploring the potential of APFs.

## Methods

**Demonstration of the general single-stage APF.** For a general 2×2 network-based lossless structure (Fig. 4a), which is represented by $S_a$ in the following text, the transmission matrix $M_0$ of the inputs and outputs can be expressed as

$$\begin{pmatrix} E_{2a} \\ E_{oa} \end{pmatrix} = \begin{pmatrix} m_{11} & m_{12} \\ m_{21} & m_{22} \end{pmatrix} \begin{pmatrix} E_{1a} \\ E_{ia} \end{pmatrix} \quad (4)$$

When the output $E_{2a}$ is connected to the input $E_{1a}$ with a lossless delay line, the relation between $E_{1a}$ and $E_{2a}$ can be expressed by $E_{1a}=E_{2a}p_0$, where $p_0=exp(i\varphi_0)$ and $\varphi_0$ is the phase shift of the delay line. The transfer function of the lossless structure can be written as

$$H_a = \frac{E_{oa}}{E_{ia}} = \frac{m_{22} - p_0 d_0^2}{1 - m_{11} p_0} \quad (5)$$

where $d_0^2$ is the determinant of $M_0$ and $d_0^2 = det(M_0)$. Then, it can be derived that

$$|H_a|^2 = \frac{m_{22} - p_0 d_0^2}{1 - m_{11} p_0} \cdot \left(\frac{m_{22} - p_0 d_0^2}{1 - m_{11} p_0}\right)^* = \frac{|d_0^2|^2 - m_{22}^* p_0 d_0^2 - m_{22} p_0^* (d_0^2)^* + |m_{22}|^2}{1 - m_{11} p_0 - m_{11}^* p_0^* + |m_{11}|^2} \quad (6)$$

In a single-stage APF, $|H_a|^2=1$ must be satisfied. For general 2×2 network-based structures, such as an all-pass MRR, the transmission matrices in the ideal lossless condition satisfy $m_{11}=m_{22}$ and $d_0^2=1$. Therefore, the transfer functions of these structures satisfy $|H_a|^2=1$, and all-pass transmission can be realized. However, there is definitely transmission loss in practice, and a loss coefficient must be added into equation (5). Therefore, equation (5) cannot be satisfied, and consequently all-pass transmission cannot be achieved. Based on this kind of structure, the method of self-compensation of loss can be used to realize a lossy APF, which is an APF with transmission loss. The method is also applicable to other structures that satisfy $|H_a|^2=1$ in lossless conditions (such as the proposed MZI-assisted MRR, in which $d_0^2 \neq 1$). The key to achieving the APF is to compensate for the transmission loss. Therefore, the lossy structure in Fig. 4b, which is represented by $S_b$ in the following text, is proposed. The transmission loss in this structure can be compensated by using a compensation, which is a power scaling copy of the input and its phase is the same as that of the input. The compensation can be easily achieved by taking a slice from the input. The two delay lines used to connect $M_1$ and $M_2$ are of the same length as that of the delay line in $S_a$, and the phase shifts of the two delay lines in $S_b$ are equal to $\varphi_0$. The amplitude transmissions of the two delay lines in $S_b$ are both $a_0$. The relation between the inputs $E_{ib}$ and $E_{1b}$ and the outputs $E_{ob}$ and $E_{2b}$ of the matrix $M_1$ and the relation between the inputs $E_c$ and $E_3$ and the outputs $E_4$ and $E_5$ of the matrix $M_2$ can be expressed by

$$\begin{pmatrix} E_{2b} \\ E_{ob} \end{pmatrix} = \begin{pmatrix} M_{11} & M_{12} \\ M_{21} & M_{22} \end{pmatrix} \begin{pmatrix} E_{1b} \\ E_{ib} \end{pmatrix}$$
$$\begin{pmatrix} E_4 \\ E_5 \end{pmatrix} = \begin{pmatrix} M_{33} & M_{34} \\ M_{43} & M_{44} \end{pmatrix} \begin{pmatrix} E_3 \\ E_c \end{pmatrix}$$
(7)

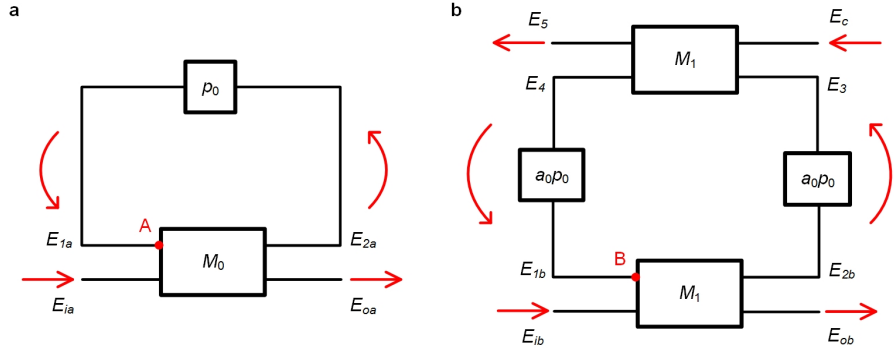

**Fig. 4.** General forms of the applicable structures for the proposed method. **a**, The ideal lossless structure of a general single-stage APF. The amplitudes of the two inputs and the two outputs of $M_0$ are represented by $E_{ia}$, $E_{1a}$, $E_{oa}$ and $E_{2a}$, respectively. $p_0=exp(i\varphi_0)$, and $\varphi_0$ is the phase shift of the delay line connecting the input $E_{1a}$ and the output $E_{2a}$. **b**, The lossy structure with self-compensation of loss. The loss in the propagation and in the transmission matrices ($M_1$ and $M_2$) are both considered in this structure. The amplitudes of the two inputs and the two outputs of $M_1$ are represented by $E_{ib}$, $E_{1b}$, $E_{ob}$ and $E_{2b}$, respectively, and the amplitudes of the two inputs and the two outputs of $M_2$ are represented by $E_3$, $E_c$, $E_4$ and $E_5$, respectively. The phase shift of the delay lines connecting $M_1$ and $M_2$ are both equal to $\varphi_0$. The amplitude transmissions of the two delay lines are both $a_0$.

The determinants of $M_1$ and $M_2$ are represented by $\det(M_1)$ and $\det(M_2)$, respectively. Considering the loss of the 2×2 network structures, we denote that $\det(M_1)=d_1^2$ ($0<d_1<1$) and $\det(M_2)=d_2^2$ ($0<d_2<1$). Based on equation (7), $E_3=E_{2b}a_0p_0$ and $E_{1b}=E_4a_0p_0$, the relations among the input $E_{ib}$, compensation $E_c$ and output $E_{ob}$ can be calculated as

$$\begin{pmatrix} E_{ib} \\ E_{ob} \end{pmatrix} = \begin{pmatrix} \dfrac{-M_{11}}{M_{12}} & \dfrac{1}{M_{12}} \\ \dfrac{-\det(M_1)}{M_{12}} & \dfrac{M_{22}}{M_{12}} \end{pmatrix} \begin{pmatrix} 0 & a_0 p_0 \\ \dfrac{1}{a_0 p_0} & 0 \end{pmatrix} \begin{pmatrix} \dfrac{-M_{44}}{M_{43}} & \dfrac{1}{M_{43}} \\ \dfrac{-\det(M_2)}{M_{43}} & \dfrac{M_{33}}{M_{43}} \end{pmatrix} \begin{pmatrix} E_c \\ E_5 \end{pmatrix} \quad (8)$$

Combining equations (6) and (8), the amplitude of the transfer function $|H_b|$ of $S_b$ can be calculated, and the APF can be realized when $|H_b|=1$. To simplify the calculations, we can conclude two simplified conditions for realizing the APF. Considering the light recirculating in the loop of $S_a$ and in the loop of $S_b$, we denote the amplitude of the optical field at a certain point, such as point A in $S_a$ and point B in $S_b$, as $E_0$. After recirculating two times in the loop of $S_a$ and returning to point A, the amplitude of the optical field changes and can be expressed as $E_0 m_{11}^2 p_0^2$. For a comparison, after recirculating one time in the loop of $S_b$ and returning to point B, the amplitude of the optical field can be expressed as $E_0 M_{11} M_{33} a_0^2 p_0^2$. Condition 1 is to ensure that the amplitude variation after two recirculations in the loop of $S_a$ is equivalent to the amplitude variation after one recirculation in the loop of $S_b$, which indicates that

$$M_{11} M_{33} a_0^2 = m_{11}^2 \quad (9)$$

To compensate for the transmission loss of the input exactly, condition 2, which requires the power ratio between the compensation and the input in $S_b$ to be equal to a specific value, also needs to be satisfied. To satisfy condition 2, the ratio between $E_c$ and $E_{ib}$ can be ascertained by a simplified method. If we assume that the number of times that the light recirculates in the loop of $S_b$ is $n$ ($n=0, 1, 2\ldots$), then the amplitude of the final output $E_{ob}$ in $S_b$ can be decomposed into a superposition of separate outputs after each recirculation of the loop, which can be expressed as $E_{ob(0)}, E_{ob(1)}, E_{ob(2)}\ldots$ and $E_{ob(n)}$ ($E_{ob(n)}$ denotes the amplitude of the output light in Fig. 4b after the $n$-th recirculation). For the light injected as the compensation, whose amplitude is $E_c$ (Fig. 4b), the output amplitude after the $n$-th recirculation in the loop can be expressed as $E_{ob(0.5+n)}$ (point B is set as the ending point of the recirculation). Similarly, in $S_a$, the amplitude of the output after the $n$-th recirculation in the loop can be expressed as $E_{oa(n)}$. Based on the relation that the loop in $S_a$ is one-half the loop length in $S_b$, to satisfy condition 2, we must ensure that for any $E_{oa(n)}$ in $S_a$, there is a corresponding $E_{ob(n/2)}$ in $S_b$ to satisfy $E_{oa(n)}=E_{ob(n/2)}$. When condition 1 is satisfied, if $E_{oa(n)}=E_{ob(n/2)}$, we can deduce that $E_{oa(n+2)}=E_{ob(n/2+1)}$. Therefore, to achieve $E_{ob}=E_{oa}$, we only need to ensure that $E_{ob(0)}=E_{oa(0)}$, $E_{ob(0.5)}=E_{oa(1)}$ and $E_{ob(1)}=E_{oa(2)}$. These relations can be calculated to be

$$E_{ib} M_{22} = E_{ia} m_{22} \quad (10)$$

$$E_c M_{12} M_{34} a_0 p_0 = E_{ia} m_{12} m_{21} p_0 \quad (11)$$

$$E_{ib} M_{12} M_{21} M_{33} a_0^2 p_0^2 = E_{ia} m_{11} m_{12} m_{21} p_0^2 \quad (12)$$

Combining equations (9)-(12), we can deduce that $M_{11}M_{22}=m_{11}m_{22}d_1^2/d_0^2$, $M_{12}M_{21}=m_{12}m_{21}d_1^2/d_0^2$ and $E_c/E_{ib}=m_{12}m_{21}M_{22}/(m_{22}M_{12}M_{34}a_0)$. In the general transmission matrices in which $m_{12}=m_{21}$ and $M_{12}=M_{21}$, we can deduce that $M_{12}=m_{21}d_1/d_0$. If we assume that the ratio between $E_c$ and $E_{ib}$ satisfies $E_c/E_{ib}=x$, it can be derived that

$$x = \frac{m_{12}M_{22}d_0}{m_{22}M_{34}d_1 a_0} \quad (13)$$

Therefore, an APF can be achieved based on the two conditions. Condition 1 ensures that the signal recirculations in the loops of the two structures are equivalent and can be expressed by equation (9). Condition 2 requires that the power ratio between the compensation and input must be equal to a specific value, which can be expressed by equation (13). By combining equations (8), (9), (13) and $M_{12}=M_{21}=m_{12}d_1/d_0=m_{21}d_1/d_0$, we can calculate that

$$H_b = \frac{E_{ob}}{E_{ib}} = \frac{d_1^2 m_{11}}{d_0^2 M_{11}} \cdot \frac{m_{22}-p_0 d_0^2}{1-m_{11}p_0} \quad (14)$$

Comparing equation (14) with equation (5), we can conclude that the transfer functions of the two structures have the same forms except an additional attenuation of $d_1^2 m_{11}/(d_0^2 M_{11})$ in $H_b$. Combining equations (6) and (14), we can also calculate that $|H_b|^2=|d_1^2 m_{11}|^2/|d_0^2 M_{11}|^2$. Obviously, $|H_b|^2$ is a constant, so we have proven that, when the two conditions are satisfied, the lossy structure (Fig. 4b) is an APF. It should be noted that to simplify the demonstration, some restrictions ($m_{12}=m_{21}$ and $M_{12}=M_{21}$) are set here and will limit the application of the method. The core of the method is to make a lossy structure equivalent to a lossless structure, in which conditions 1 and 2 must be satisfied simultaneously. In particular cases, some restrictions can be changed accordingly while the core of the method remains the same.

**Demonstration of the APF based on the add-drop MRR.** We can find some examples for a single-stage lossy APF. A possible structure is based on an add-drop MRR. Because the coupling loss also affects the transmission, here, the effect of the coupling loss on the transmission of the lossy APF will be discussed separately. In the add-drop MRR (Fig. 1b), we assume that the self-coupling and cross-coupling coefficients with coupling loss are $R_1$ and $K_1$, respectively. For the structures shown in Figs 1c and b, equations (4) and (7) can be rewritten as

$$\begin{pmatrix} E_{2a} \\ E_{oa} \end{pmatrix} = \begin{pmatrix} r_1 & ik_1 \\ ik_1 & r_1 \end{pmatrix} \begin{pmatrix} E_{1a} \\ E_{ia} \end{pmatrix} \quad (15)$$

and

$$\begin{pmatrix} E_{2b} \\ E_{ob} \end{pmatrix} = \begin{pmatrix} R_1 & iK_1 \\ iK_1 & R_1 \end{pmatrix} \begin{pmatrix} E_{1b} \\ E_{ib} \end{pmatrix}$$
$$\begin{pmatrix} E_4 \\ E_5 \end{pmatrix} = \begin{pmatrix} R_2 & iK_2 \\ iK_2 & R_2 \end{pmatrix} \begin{pmatrix} E_3 \\ E_c \end{pmatrix} \quad (16)$$

respectively. Here, $\det(M_0)=r_1^2+k_1^2=d_0^2=1$, $\det(M_1)=R_1^2+K_1^2=d_1^2$ and the coupling loss in coupler $C_1$ is $1-d_1^2$. From $M_{11}M_{22}=m_{11}m_{22}d_1^2/d_0^2$ and $M_{12}M_{21}=m_{12}m_{21}d_1^2/d_0^2$, we can obtain $R_1^2=r_1^2d_1^2$ and $K_1^2=k_1^2d_1^2$. Therefore, from equation (9), we can deduce that

$$R_1 R_2 a_0^2 = r_1^2 = R_1^2 / d_1^2 \tag{17}$$

From equation (17), we can obtain $R_1=R_2 d_1^2 a_0^2$. Based on equation (13), $R_1^2=r_1^2 d_1^2$ and $K_1^2=k_1^2 d_1^2$, $x$ can be derived as

$$x = \frac{k_1 R_1}{r_1 K_2 d_1 a_0} = \frac{K_1}{K_2 d_1 a_0} \tag{18}$$

Compared with the occasion when the coupling loss in coupler $C_1$ is not separated from the round-trip amplitude transmission $a$, where $a=d_1^2 a_0^2$, the outcomes of $x$ in the two cases are equivalent. When the coupling loss is separated from the round-trip amplitude transmission, the only alteration of the transfer function is that a modification should be added to equation (2). Based on equation (14) and $R_1^2=r_1^2 d_1^2$, the transfer function of the add-drop MRR can be derived as

$$H = d_1 \frac{r_1 - p_0}{1 - r_1 p_0} = d_1 \frac{r_1 - e^{i(\varphi/2)}}{1 - r_1 e^{i(\varphi/2)}} \tag{19}$$

where $\varphi_0=\varphi/2$. From equations (6) and (19), we can calculate that $|H|^2=d_1^2$, which indicates that the lossy structure based on the add-drop MRR is an APF after self-compensation of loss. Therefore, the general method can be applied to the demonstration of an APF based on the add-drop MRR. We also demonstrated that the coupling loss does not influence the feasibility of the APF. Therefore, the coupling loss can be included in the amplitude transmission of the structure to simplify the demonstration.

**Demonstration of the APF based on an MZI-assisted MRR.** To simplify the demonstration of the MZI-based APF, here the coupling loss will not be discussed separately. For the structures shown in Figs 2a and b, equations (4) and (7) can be rewritten as

$$\begin{pmatrix} E_{2a} \\ E_{oa} \end{pmatrix} = \begin{pmatrix} p_1(r_1^2 - k_1^2) & -2ir_1 k_1 p_1 \\ -2ir_1 k_1 p_1 & p_1(r_1^2 - k_1^2) \end{pmatrix} \begin{pmatrix} E_{1a} \\ E_{ia} \end{pmatrix} \tag{20}$$

and

$$\begin{pmatrix} E_{2b} \\ E_{ob} \end{pmatrix} = \begin{pmatrix} a_1 p_1(r_1^2 - k_1^2) & -2ia_1 r_1 k_1 p_1 \\ -2ia_1 r_1 k_1 p_1 & a_1 p_1(r_1^2 - k_1^2) \end{pmatrix} \begin{pmatrix} E_{1b} \\ E_{ib} \end{pmatrix}$$

$$\begin{pmatrix} E_4 \\ E_5 \end{pmatrix} = \begin{pmatrix} a_1(p_1 r_2^2 - p_2 k_2^2) & -ia_1 r_2 k_2(p_1 + p_2) \\ -ia_1 r_2 k_2(p_1 + p_2) & a_1(p_2 r_2^2 - p_1 k_2^2) \end{pmatrix} \begin{pmatrix} E_3 \\ E_c \end{pmatrix} \tag{21}$$

where $p_1=exp(i\varphi_1)$ and $p_2=exp(i\varphi_2)$.

From equation (20), we can deduce that the determinant $det(M_0)=d_0^2=p_1^2$. Based on equation (20) and $E_{1a}=E_{2a}p_0$, the transfer function of the ideal lossless structure shown in Fig. 2b can be calculated as

$$H_a = \frac{p_1(r_1^2 - k_1^2) - p_0 p_1^2}{1 - p_0 p_1(r_1^2 - k_1^2)} \quad (22)$$

Combining equations (6) and (22), we can derive that $|H_a|^2=1$, which indicates that the ideal lossless structure (Fig. 2b) is an APF. It can also be calculated that $\det(M_1)=d_1^2=a_1^2 p_1^2$. From equations (9), (20) and (21), we can derive that

$$a_0^2 a_1^2 (p_1 r_2^2 - p_2 k_2^2) = p_1(r_1^2 - k_1^2) \quad (23)$$

Combining equations (13), (20) and (21), we can derive that

$$x = \frac{2 r_1 k_1 p_1}{r_2 k_2 (p_1 + p_2) a_0 a_1} \quad (24)$$

By substituting $p_1=exp(i\varphi_1)$, $p_2=exp(i\varphi_2)$, $y=exp[i(\varphi_2-\varphi_1)]$ and $r_2^2+k_2^2=1$ into equations (23) and (24), we can derive equation (3). Combining equations (3), (21), $E_3=E_{2b}a_0 p_0$ and $E_{1b}=E_4 a_0 p_0$, we can also derive the transfer function of the MZI-assisted MRR

$$H_b = a_1 \frac{p_1(r_1^2 - k_1^2) - p_0 p_1^2}{1 - p_0 p_1(r_1^2 - k_1^2)} \quad (25)$$

From equations (6) and (25), we can deduce that $|H_b|^2=a_1^2$, which proves that the proposed MZI-assisted MRR is an APF. Comparing equations (22) and (25), we can observe that the transfer function of the APF based on the MZI-assisted MRR (Fig. 2a) is equal to that of the ideal lossless APF (Fig. 2b), except for the additional attenuation $a_1$ in equation (25). The additional loss is generated by the lossy waveguides of $MZI_1$ and $MZI_2$ (Fig. 2a), and the waveguides are lossless in the structure in Fig. 2b. Neglecting the additional attenuation $a_1$, here we have demonstrated that the two APFs in Figs 2a and b are equivalent to each other.

**Adjusting the fabricated device to function as an equivalent optical waveguide.** In Fig. 2c, by adjusting the voltage applied to the micro-heater deposited on the lower arm of $MZI_0$ to change the phase difference between the two arms of $MZI_0$, the power splitting ratio of $MZI_0$ can be adjusted as well. At first, we adjust the power splitting ratio of $MZI_0$ to maximize the power of the optical signal output from $GC_2$, which indicates that no optical power splits into the upper arm of the output waveguides of $MZI_0$. Then, we adjust the voltage applied to the micro-heater on the lower arm of $MZI_1$ to eliminate the notches at the resonance wavelengths of the ring, which indicates that the optical signal in the lower arm of the output waveguides of $MZI_0$ is directly coupled to $GC_2$ and no optical power is coupled into the ring. Consequently, the device works as a single-mode waveguide.

**Device fabrication and experimental details.** The device is fabricated on a standard SOI wafer with a 220-nm-thick silicon layer on a 2-μm-thick buried oxide layer. All the silicon waveguides are strip waveguides, and the width and depth of the waveguide are 500 nm and 220 nm, respectively. After the 220-nm silicon etch process, the silicon waveguides are then covered by $SiO_2$ deposition with a thickness of 1.2 μm. The related metal layer is deposited upon the SiO$_2$ layer to form the micro-heater.

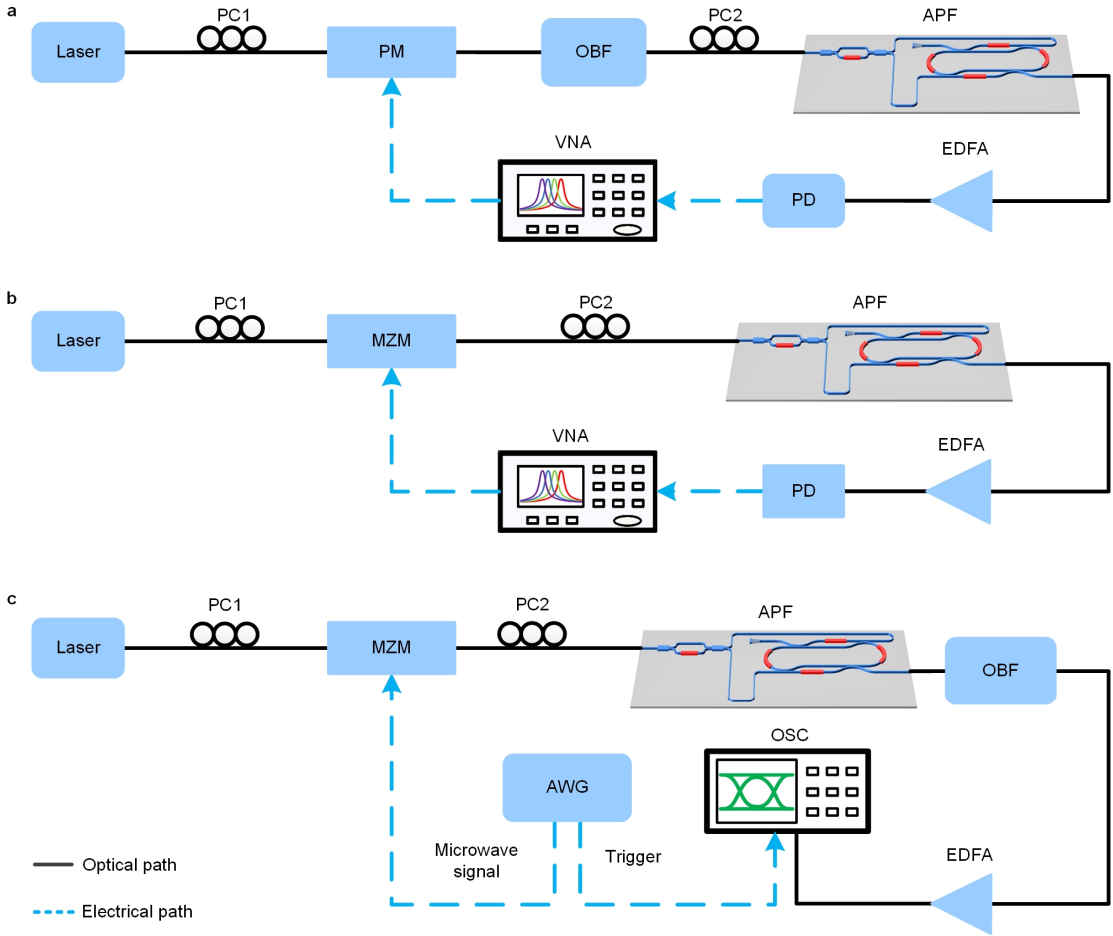

**Fig. 5.** Experimental setups. **a**, **b**, **c**, Setups for measuring the amplitude and phase responses of the fabricated APF (**a**), implementing a microwave photonic notch filter (**b**) and measuring the time delay of the APF (**c**). PC, polarization controller; PM, phase modulator; OBF, optical bandpass filter; EDFA, erbium-doped fibre amplifier; PD, photodetector; VNA, vector network analyzer; MZM, Mach-Zehnder modulator; AWG, arbitrary waveform generator; OSC, oscilloscope. It should be noted that the light is coupled into and out of the photonic chip by using a tapered and lensed fibre and the grating couplers in the fabricated device.

The experimental setup for measuring the amplitude and phase responses of the fabricated APF is shown in Fig. 5a. The continuous-wave light emitted from a laser source (Koheras BASIK) is injected into a phase modulator (PM, Covega LN66S-FC). The optical carrier is then modulated by microwave signals emitted by the VNA (Anritsu MS4647B). An optical bandpass filter (OBF) is used to realize single-sideband modulation after PM, and the response of the APF will be mapped from the optical domain to the electrical domain after the photodetector (PD, SHF AG Berlin). The frequency responses from 0 to 40 GHz are measured by VNA, and the range of the measurement is limited by the bandwidth of the PM. The demonstration of the microwave photonic phase shifter[6,31-33] is also based on this experimental setup. By setting the optical carrier wavelength within the phase shift region of the APF, a microwave phase shift can be achieved by beating the sideband and the optical carrier in the PD. By adjusting the micro-heater deposited on the MRR (Fig. 2c), the phase offset between the optical carrier and the sideband is consequently changed. Therefore, the phase shift of the microwave is changed and measured by the VNA.

Based on the APF, a microwave photonic notch filter can be implemented[34,35], and the experimental setup is shown in Fig. 5b. A Mach-Zehnder modulator (MZM, Fujitsu FTM7938EZ) is used to implement the amplitude modulation. The APF works as an optical phase shifter to change the phase of one sideband, and the phase difference between the two beat frequencies, which are generated by beating the optical carrier and the ±1st order sidebands, is also changed consequently. When the phase difference between the two beat frequencies is π, a notch band with a high rejection ratio is generated because the two beat frequencies with the same amplitude and counter phases cancel each other out. Tuning the phase shift region of the APF by the micro-heater deposited on the MRR, the central frequency of the notch response can be tuned from 0 to 40 GHz with a rejection ratio of more than 50 dB.

The experimental setup for measuring the time delay based on the APF[36,37] is shown in Fig. 5c. The microwave signals applied to the MZM are generated from an arbitrary waveform generator (AWG, Keysight M8195A). The modulated signals are injected into the APF. By setting the optical spectrum of the modulated optical signal within the phase shift region of the APF, a time delay can be achieved. Here, the OBF is used to suppress the noise and improve the signal-to-noise ratio of the signal. Another low-frequency sinusoidal signal generated by the AWG is used as the frequency clock to trigger the oscilloscope. By adjusting the phase shift region of the APF, different time delays can be obtained, and the measured maximum delay time is approximately 200 ps.

**Data availability**.
The data that support the findings of this study are available from the corresponding authors upon request.

**References**

1      Magden, E. S. et al. Transmissive silicon photonic dichroic filters with spectrally selective waveguides. *Nat. Commun.* **9**, 3009 (2018).
2      Sancho, J. et al. Integrable microwave filter based on a photonic crystal delay line. *Nat. Commun.* **3**, 1075 (2012).
3      Fandiño, J. S., Muñoz, P., Doménech, D. & Capmany, J. A monolithic integrated photonic microwave filter. *Nat. Photon.* **11**, 124 (2016).
4      Supradeepa, V. R. et al. Comb-based radiofrequency photonic filters with rapid tunability and high selectivity. *Nat. Photon.* **6**, 186 (2012).
5      Zhuang, L., Roeloffzen, C. G. H., Hoekman, M., Boller, K.-J. & Lowery, A. J. Programmable photonic signal processor chip for radiofrequency applications. *Optica* **2**, 854-859 (2015).
6      Adams, D. B. & Madsen, C. K. A novel broadband photonic RF phase shifter. *J.*

|   |   |
|---|---|
|   | *Lightwave Technol.* **26**, 2712-2717 (2008). |
| 7 | Yang, W. et al. High speed optical phased array using high contrast grating all-pass filters. *Opt. Express* **22**, 20038-20044 (2014). |
| 8 | Madsen, C. K. & Lenz, G. Optical all-pass filters for phase response design with applications for dispersion compensation. *IEEE Photonics Technol. Lett.* **10**, 994-996 (1998). |
| 9 | Jablonski, M., Takushima, Y. & Kikuchi, K. The realization of all-pass filters for third-order dispersion compensation in ultrafast optical fiber transmission systems. *J. Lightwave Technol.* **19**, 1194 (2001). |
| 10 | Lunardi, L. M. et al. Tunable dispersion compensation at 40-Gb/s using a multicavity Etalon all-pass filter with NRZ, RZ, and CS-RZ modulation. *J. Lightwave Technol.* **20**, 2136 (2002). |
| 11 | Madsen, C. K. Subband all-pass filter architectures with applications to dispersion and dispersion-slope compensation and continuously variable delay lines. *J. Lightwave Technol.* **21**, 2412 (2003). |
| 12 | Kim, J., Ko, Y., Kim, H. & Chung, Y. in *2013 18th OptoElectronics and Communications Conference held jointly with 2013 International Conference on Photonics in Switching.*   TuPL_11 (Optical Society of America). |
| 13 | Maram, R., Kong, D., Galili, M., Oxenløwe, L. K. & Azaña, J. Ultrafast all-optical clock recovery based on phase-only linear optical filtering. *Opt. Lett.* **39**, 2815-2818 (2014). |
| 14 | Zhuang, L. et al. Novel microwave photonic fractional Hilbert transformer using a ring resonator-based optical all-pass filter. *Opt. Express* **20**, 26499-26510 (2012). |
| 15 | Liu, W. et al. A fully reconfigurable photonic integrated signal processor. *Nat. Photon.* **10**, 190 (2016). |
| 16 | Lenz, G. & Madsen, C. K. General optical all-pass filter structures for dispersion control in WDM systems. *J. Lightwave Technol.* **17**, 1248-1254 (1999). |
| 17 | Ibrahim, M. A., Kuntman, H. & Cicekoglu, O. First-order all-pass filter canonical in the number of resistors and capacitors employing a single DDCC. *Circuits Syst. Signal Process* **22**, 525-536 (2003). |
| 18 | Minaei, S. & Yuce, E. Novel voltage-mode all-pass filter based on using DVCCs. *Circuits Syst. Signal Process* **29**, 391-402 (2010). |
| 19 | Minaei, S. & Cicekoglu, O. A resistorless realization of the first-order all-pass filter. *Int. J. Electron.* **93**, 177-183 (2006). |
| 20 | Gupta, S., Zhang, Q., Zou, L., Jiang, L. J. & Caloz, C. Generalized coupled-line all-pass phasers. *IEEE Trans. Microw. Theory Tech.* **63**, 1007-1018 (2015). |
| 21 | Guyette, A. C., Naglich, E. J. & Shin, S. Switched allpass-to-bandstop absorptive filters with constant group delay. *IEEE Trans. Microw. Theory Tech.* **64**, 2590-2595 (2016). |
| 22 | Naglich, E. J., Lee, J., Peroulis, D. & Chappell, W. J. Switchless tunable bandstop-to-all-pass reconfigurable filter. *IEEE Trans. Microw. Theory Tech.* **60**, 1258-1265 (2012). |
| 23 | Viveiros, D., Consonni, D. & Jastrzebski, A. K. A tunable all-pass MMIC active |

phase shifter. *IEEE Trans. Microw. Theory Tech.* **50**, 1885-1889 (2002).

24  Bogaerts, W. et al. Silicon microring resonators. *Laser Photon. Rev.* **6**, 47-73 (2012).

25  Zhou, L. & Poon, A. W. Electrically reconfigurable silicon microring resonator-based filter with waveguide-coupled feedback. *Opt. Express* **15**, 9194-9204 (2007).

26  Thomson, D. J., Hu, Y., Reed, G. T. & Fedeli, J. Low loss MMI couplers for high performance MZI modulators. *IEEE Photonics Technol. Lett.* **22**, 1485-1487 (2010).

27  Feng, D. J. Y. & Lay, T. S. Compact multimode interference couplers with arbitrary power splitting ratio. *Opt. Express* **16**, 7175-7180 (2008).

28  Dong, P. et al. Thermally tunable silicon racetrack resonators with ultralow tuning power. *Opt. Express* **18**, 20298-20304 (2010).

29  Dong, J. et al. Photonic Hilbert transformer employing on-chip photonic crystal nanocavity. *J. Lightwave Technol.* **32**, 3704-3709 (2014).

30  Nguyen, T. G. et al. Integrated frequency comb source based Hilbert transformer for wideband microwave photonic phase analysis. *Opt. Express* **23**, 22087-22097 (2015).

31  Wang, X., Chan, E. H. W. & Minasian, R. A. All-optical photonic microwave phase shifter based on an optical filter with a nonlinear phase response. *J. Lightwave Technol.* **31**, 3323-3330 (2013).

32  Pu, M. et al. Widely tunable microwave phase shifter based on silicon-on-insulator dual-microring resonator. *Opt. Express* **18**, 6172-6182 (2010).

33  Chang, Q. et al. A tunable broadband photonic RF phase shifter based on a silicon microring resonator. *IEEE Photonics Technol. Lett.* **21**, 60-62 (2009).

34  Yan, Y. & Yao, J. A tunable photonic microwave filter with a complex coefficient using an optical RF phase shifter. *IEEE Photonics Technol. Lett.* **19**, 1472-1474 (2007).

35  Zhang, Y. & Pan, S. Complex coefficient microwave photonic filter using a polarization-modulator-based phase shifter. *IEEE Photonics Technol. Lett.* **25**, 187-189 (2013).

36  Liu, Y., Choudhary, A., Marpaung, D. & Eggleton, B. J. Gigahertz optical tuning of an on-chip radio frequency photonic delay line. *Optica* **4**, 418-423 (2017).

37  Wang, X. et al. Continuously tunable ultra-thin silicon waveguide optical delay line. *Optica* **4**, 507-515 (2017).

**Acknowledgements** This work was supported in part by the National Natural Science Foundation of China (61501194, 11664009), the Hubei Provincial Natural Science Foundation of China (2015CFB231, 2016CFB370), and the Fundamental Research Funds for the Central Universities (HUST: 2016YXMS025). We would also like to thank Prof. Jianping Yao of University of Ottawa for discussions.

**Author contributions** L.X., Y.Y. and X.L. conceived the idea. L.X. and Y.Y.